# About the connection of the electron binding energy of a single carbon anion with binding energies of an electron attached to carbon molecules

A. S. Baltenkov[1] and I. Woiciechowski[2]

[1]*Arifov Institute of Ion-Plasma and Laser Technologies,*
*100125, Tashkent, Uzbekistan*
[2]*Alderson Broaddus University, 101 College Hill Drive, Philippi, 26416, WV, USA*

**Abstract**
*We demonstrate that the model of zero-range potentials can be successfully employed for the description of attached electrons in atomic and molecular anions, for example, negatively charged carbon clusters. To illustrate the capability of the model we calculate the energies of the attached electron for the family of carbon cluster anions consisting of two-, three- (equilateral triangle), and four (tetrahedron) carbon atoms equidistant from each other as well as for a $C_3$ molecule having a chain structure. The considered approach can be easily extended to carbon clusters containing an arbitrary number of atoms arranged in an arbitrary configuration.*

One of the tasks of quantum chemistry is determining the energy levels of an electron in a potential field, which is a superposition of centrally symmetric potentials with centers at the positions of atoms in a molecule. Solutions to quantum mechanical problems of this type are extremely difficult. Reliable results of ab initio calculations are only available for clusters with relatively small numbers of constituents. For example, the structure- and energy calculations of carbon cluster anions $C_N^-$ are reported for only up to $N = 18$ in refs [1, 2]. There exists, however, such a field, for which the problem of electron motion has a simple analytical solution. This is the field formed by zero-range potentials. Within the approach described in [3, 4], the molecular constituents are replaced by a set of $N$ zero-range potential wells located at the positions $\mathbf{R}_j$ ($j =1, 2, 3…N$) of the nuclei. The goal of the present communication is a demonstration that the model of zero-range potentials [3, 4] can be successfully employed for the description of attached electrons in atomic and molecular anions, for example, negatively charged carbon clusters with arbitrary number of constituents arranged in an arbitrary configuration. We briefly outline the model in the next paragraph.

The wave function of the electron $\psi(\mathbf{r})$ in a multicenter field, being a solution of the Schrödinger equation everywhere in space except the points $\mathbf{R}_j$ and decreasing at $r \to \infty$, is a superposition of the wave functions of single carbon atoms

$$\psi(\mathbf{r}) = \sum_{j=1}^{N} c_j \frac{e^{-\kappa r_j}}{r_j}, \qquad \mathbf{r}_j = \mathbf{r} - \mathbf{R}_j, \qquad (1)$$

where $\kappa$ is the magnitude of the wave vector of the cluster. At the positions $\mathbf{r} = \mathbf{R}_j$, function (1) has to satisfy the boundary conditions

$$\psi\big|_{|\mathbf{r}-\mathbf{R}_j|\to 0} = b_j [\frac{1}{|\mathbf{r}-\mathbf{R}_j|} - \alpha], \qquad (2)$$

where $\alpha = \sqrt{2E_0}$, $E_0$ is the binding energy of the electron in a single zero-range potential well. We use the atomic units throughout the paper. Application of the boundary conditions (2) to the wave function (1) results in a system of homogeneous linear equations for the coefficients $c_j$ in equation (1)

$$(\alpha - \kappa)c_i + \sum_{\substack{j=1 \\ j \neq i}}^{N} c_j \frac{e^{-\kappa R_{ij}}}{R_{ij}} = 0, \quad i=1, 2, 3...N, \quad \mathbf{R}_{ij} = \mathbf{R}_i - \mathbf{R}_j. \tag{3}$$

The system of equations (3) has nontrivial solutions when its determinant, being a function of κ,

$$W(\kappa) = 0. \tag{4}$$

Equation (4) is a transcendent equation connecting the wave vector $\kappa$ with the wave vector $\alpha$. It has several solutions that correspond to discrete energy levels ($\kappa>0$). The number of energy levels does not exceed the number of potential wells $N$. Other states of the negative ion belong to the continuum. The energies of the bound states of the molecular electron $E_N = -\kappa^2/2$.

To illustrate the capability of the above formulas, let us consider a family of carbon clusters consisting of two-, three- (equilateral triangle) and four (tetrahedron) carbon atoms equidistant from each other. These model carbon structures exhaust all possible configurations, in which all center-to-center distances are the same and equal to $R$. In terms of the notation $A = e^{-\kappa R}/R$ and $B = \alpha - \kappa$, the determinant $W$ of the system of two equations reads

$$W_2(\kappa) = \begin{Vmatrix} B & A \\ A & B \end{Vmatrix} = (B^2 - A^2). \tag{5}$$

For the system of 3 equations the determinant is

$$W_3(\kappa) = \begin{Vmatrix} B & A & A \\ A & B & A \\ A & A & B \end{Vmatrix} = (A-B)^2(2A+B). \tag{6}$$

The system of 4 equations leads to the determinant

$$W_4(\kappa) = \begin{Vmatrix} B & A & A & A \\ A & B & A & A \\ A & A & B & A \\ A & A & A & B \end{Vmatrix} = (A-B)^3(3A+B). \tag{7}$$

Evaluating the determinants (5) - (7), we use the inter-atomic distance $R = 2.751$ atomic units and $\alpha = \sqrt{1.27/27.2} = 0.3055$ a.u. (here 1.27 eV is the electron affinity of C⁻). The graphs of the determinants versus the wave numbers $\kappa$ are presented in Figure 1. The values of $\kappa$ at which the curves $W_i(\kappa)$ cross the X axis are collected in Table 1 along with the corresponding energy levels $E_N$.

**Table 1**. Zeros of the determinant of the system of equations (3).

| Determinant | Zeros of determinants, a.u. | Energies $E_N = -\kappa^2/2$, eV |
|---|---|---|
| $W_2(\kappa)$ | $\kappa$=0.42 | -2.40 eV |
| $W_3(\kappa)$ | $\kappa$=0.49 | -3.26 eV |
| $W_4(\kappa)$ | $\kappa$=0.55 | -4.11 eV |
|  |  |  |
| $W_3(\kappa)$ chain | $\kappa$=0.26 | -0.919 eV |

| $W_3(\kappa)$ chain | $\kappa=0.46$ | -2.878 eV |

Along with the equilateral triangular configuration, one can consider a linear chain of three carbon atoms with the distance $R/2$ between the closest centers. The determinant corresponding to tpy configuration of atoms has the form [5]

$$W_3(\kappa)\text{chain} = \begin{Vmatrix} B & C & A \\ C & B & C \\ A & C & B \end{Vmatrix} = (A-B)(B^2 + AB - 2C^2), \qquad (8)$$

where $A$ and $B$ are the same as for the previous determinants, $C = 2e^{-\kappa R/2}/R$. Zeroes of epy determinant (8) are shown in two lower lines in Table 1 and in Figure 1. New energy levels at $\kappa = 0.26$ a.u. and at $\kappa = 0.46$ a.u. appear for the chain structure. As we can see, the number of energy levels and their positions are defined by the geometry of the molecule constituents.

**Conclusion**

In atomic physics, the simplest system that can be modeled by a particle in the field of the zero-range wells is a negative ion of an atom with an external weakly bounded electron in the s-state. Most of the time the outer electron spends outside the potential well, where it can be considered free. The zero-range potential approach opens perspectives for a study of the electron structure of more complex carbon structures such as fullerenes and their derivatives. The wave functions and energies of the attached electron were calculated earlier in [6] for single-cage $(C_N)^-$ and multi-cage (onion-like) $(C_M@C_N@...)^-$ singly charged fullerene anions using the model of spherical potential well. This approach is only applicable for spherically symmetric structures. The model described here can be used for carbon structures of arbitrary geometries. The input information for such calculations includes the coordinates of carbon atoms (for example in $C_{20}$ [7]) and the binding energy of a single carbon anion. The determinant of the system of equations (3) can be calculated numerically using standard software packages such as MATLAB [8].

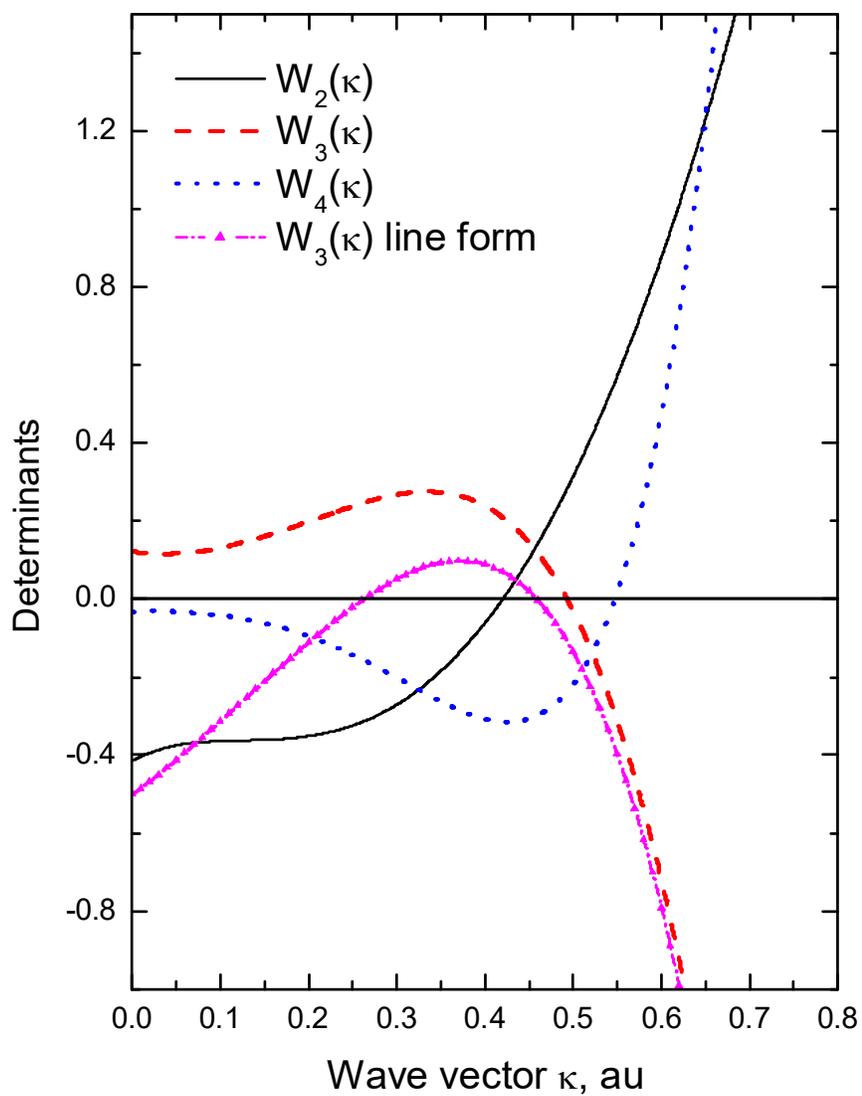

**Figure** 1. Determinants of the system of equations (3) vs the wave vector $\kappa$ for carbon molecules $C_2$, $C_3$, $C_4$.